\documentclass[%
 reprint,
 amsmath,amssymb,
 aps,superscriptaddress
]{revtex4-2}

\usepackage{graphicx}
\usepackage{dcolumn}
\usepackage{bm}
\usepackage{tikz,xcolor}
\usepackage{physics}
\usepackage{float}
\usepackage{bibunits}
\usepackage{multirow}
\usepackage{hyperref}
\usepackage{wrapfig}
\usepackage[normalem]{ulem}
\usepackage[version=4]{mhchem}
\usepackage{bm}
\usepackage{braket}
\usepackage{orcidlink}

\newcommand{\editorr}[2]{%
  \expandafter\newcommand\csname #1note\endcsname[1]{%
    \textcolor{#2}{(\textbf{#1:} ##1)}}%
  \expandafter\newcommand\csname #1\endcsname[1]{%
    \textcolor{#2}{##1}}%
  \expandafter\newcommand\csname #1cancel\endcsname[1]{%
    \textcolor{#2}{\sout{##1}}}%
  \expandafter\newcommand\csname #1change\endcsname[2]{%
    \textcolor{#2}{\sout{##1} ##2}}%
  \newenvironment{#1text}{\color{#2}}{\color{black}}
}

\editorr{LB}{red}

\makeatletter
\def\maketitle{
\@author@finish
\title@column\titleblock@produce
\suppressfloats[t]}
\makeatother

\begin{document}
\begin{bibunit}[apsrev4-2]

\title{Resonant Raman spectroscopies beyond density-functional theory}

\author{Aleksandr Poliukhin\orcidlink{0000-0001-8519-872X}}
\email{aleksandr.poliukhin@epfl.ch}
\affiliation{Theory and Simulation of Materials (THEOS), École polytechnique fédérale de Lausanne, 1015 Lausanne, Switzerland}
\author{Corto Babs Aubry}%
\affiliation{Theory and Simulation of Materials (THEOS), École polytechnique fédérale de Lausanne, 1015 Lausanne, Switzerland}
\author{Lorenzo Bastonero\orcidlink{0000-0001-9374-1876}}
\affiliation{U Bremen Excellence Chair, Bremen Center for Computational Materials Science, and MAPEX Center for Materials and Processes, University of Bremen, D-28359 Bremen, Germany}
\author{Nicola Marzari\orcidlink{0000-0002-9764-0199}}
\affiliation{Theory and Simulation of Materials (THEOS), École polytechnique fédérale de Lausanne, 1015 Lausanne, Switzerland}
\affiliation{U Bremen Excellence Chair, Bremen Center for Computational Materials Science, and MAPEX Center for Materials and Processes, University of Bremen, D-28359 Bremen, Germany}
\affiliation{PSI Center for Scientific Computing, Theory and Data, 5232 Villigen PSI, Switzerland}%
\date{\today}
\begin{abstract}
Resonant Raman spectroscopy probes, in a single measurement, how electrons and phonons couple in a material.
Density-functional theory (DFT) typically reproduces well phonon frequencies, but resonant Raman intensities hinge on electron-phonon matrix elements and electronic transitions that are far more sensitive to the underlying exchange-correlation approximation.
However, electron-phonon coupling has so far been accessible only through linear-response theories developed for a handful of semilocal DFT methods, leaving the sensitivity of resonant Raman intensities to the electronic-structure approximation essentially unexplored.
Here, we introduce a general finite-difference framework that can compute resonant Raman tensors for any electronic-structure method capable of delivering forces, eigenvalues, and wavefunctions of pristine and displaced configurations.
We apply the formalism to graphene and monolayer \ce{MoS2}, using hybrid functionals or meta-GGAs, and show that these approaches systematically enhance electron-phonon couplings relative to semilocal DFT, reflecting reduced dielectric overscreening. 
A decomposition of the Raman tensor shows that accurate intensities require electronic eigenvalues and electron-phonon matrix elements to be treated consistently at the same level of theory.
Among the approaches tested, hybrid functionals provide the best overall agreement with experiment.
Because the framework needs only quantities every electronic-structure code already produces, it opens the door to systematic, beyond-DFT Raman characterization or benchmarking against experiments, especially for 2D materials.
\end{abstract}
\maketitle

Resonant Raman spectroscopy is one of the most informative techniques, able to probe the coupling between electrons and phonons in a material in a single measurement.
Thanks to its sensitivity to the laser frequency, this technique is particularly suited to materials whose spontaneous Raman cross-sections are too small for practical characterizations, such as atomically thin transition-metal dichalcogenides like \ce{MoS2}~\cite{placidi_multiwavelength_2015}, where tuning the laser into resonance with an electronic transition enhances the signal by orders of magnitude.
A multiwavelength excitation setup can even discriminate Raman modes arising from secondary phases in thin films~\cite{Dimitrievska2014}.

The challenge with this technique resides in its interpretation, a task that can be highly non-trivial, even more so when resonances are involved.
Theoretical calculations, typically based on density-functional theory (DFT), have been fundamental in the interpretation of the resonant Raman spectra of many materials, from multi-layer graphene~\cite{Ferrari2006, Venezuela2011} to entire reference databases of resonant Raman spectra for 2D materials~\cite{Taghizadeh2020}.
Nevertheless, the complexity of these calculations made their application scarce for a number of reasons: (i) the lack of specialized, automated, and user-friendly codes; (ii) the need for dense integration grids, and hence for sophisticated interpolation schemes, in the calculation of the Raman cross-section; (iii) the restriction of most implementations to DFT and its (semi)local approximation of the exchange-correlation (XC) functional.
The first two points are being addressed by an increasing number of codes supporting resonant Raman calculations, such as \texttt{EPIQ}~\cite{Marini2024} and \texttt{QR$^2$}~\cite{huang_qr2-code_2025} interfaced with the \texttt{EPW} package~\cite{lee_electronphonon_2023}, using smart interpolation schemes~\cite{Giustino_2007} based on Wannier functions~\cite{marzari_maximally_1997, marzari_maximally_2012}.
The third point, however, remains largely unaddressed.
For this reason, how resonant Raman intensities respond to the electronic-structure method itself is still essentially unexplored.
This is highly relevant because semilocal DFT, while typically reliable for phonon frequencies, can often misrepresent the electronic structure and, by extension, the electron-phonon couplings that enters the resonant cross-section, leading to poor predictions of the mode intensities.
For instance, electron-phonon couplings~\cite{zhou_abinitio_2021} and carrier mobilities~\cite{poliukhin_carrier_2025} have already been demonstrated to be very sensitive to the underlying electronic-structure method.

Here we close this gap by proposing a general finite-difference framework for the calculation of resonant Raman intensities, complementary to the one proposed in Ref.~\citenum{bastonero_automated_2024} for non-resonant Raman spectra.
The approach uses only the forces, eigenvalues, and wavefunctions of pristine and displaced configurations, and is therefore applicable to any electronic-structure method able to deliver these quantities: hybrid~\cite{becke_new_1993,Heyd2003} and meta-GGA functionals~\cite{Furness2020}, many-body perturbation theory~\cite{gw_1985,onida_electronic_2002,Golze2019}, Koopmans-compliant functionals~\cite{dabo_koopmans_2010,borghi_koopmans-compliant_2014,nguyen_koopmans-compliant_2018,colonna_koopmans-compliant_2019}, and beyond.
In the following, we start by briefly describing the framework and its workflow.
We then apply it to calculate the resonant Raman spectra of graphene and \ce{MoS2} monolayers with different electronic structure methods, compare the results to experiments, and finally investigate the interplay between electronic bands, phonon frequencies, and electron-phonon couplings.
%
\begin{figure*}
    \centering
    \includegraphics[width=\textwidth]{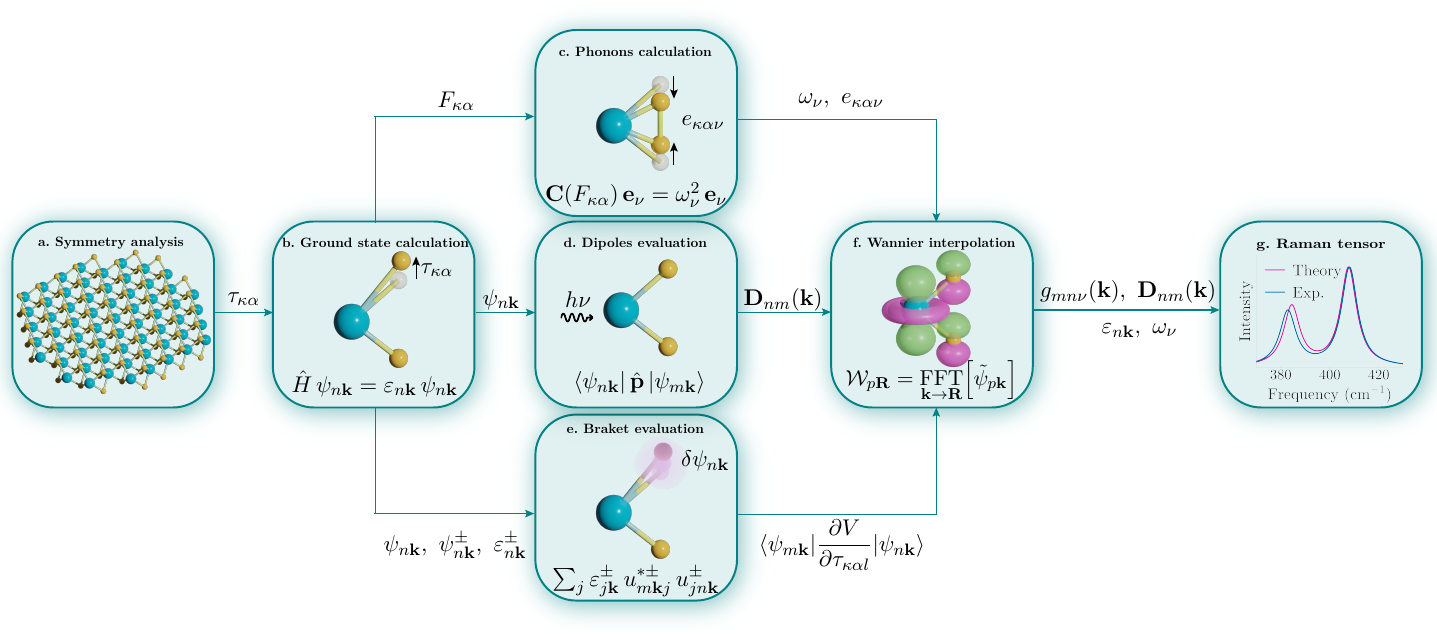}
    \caption{Schematic representation of the computational workflow for resonant Raman spectra. (a) A symmetry analysis first identifies the irreducible atomic displacements $\tau_{\kappa\alpha}$. (b) Ground-state calculations are then performed for the pristine and displaced configurations, providing the forces $F_{\kappa\alpha}$, the wavefunctions $\psi_{n\textbf{k}}$, $\psi^{\pm}_{n\textbf{k}}$, and the eigenvalues $\varepsilon_{n\textbf{k}}$, $\varepsilon^{\pm}_{n\textbf{k}}$. After that, three independent steps follow: (c) the phonon frequencies $\omega_{\nu}$ and eigenvectors $e_{\kappa\alpha\nu}$ are obtained by diagonalizing the dynamical matrix $\mathbf{C}(F_{\kappa\alpha})$ constructed from the forces; (d) the optical dipole matrix elements $\mathbf{D}_{nm}(\mathbf{k})$ are evaluated from the pristine wavefunctions; and (e) the electron-phonon matrix elements of Eq.~\eqref{eq:dVdt} are computed from the overlaps between pristine and displaced wavefunctions, following Eq.~\eqref{eq:projectability}. (f) All quantities are then interpolated on a fine $\textbf{k}$ grid using Wannier functions. (g) Finally, the Raman tensor is assembled according to Eq.~\eqref{eq:raman_tensor} and the spectrum is computed via Eq.~\eqref{eq:raman_main}.
    }
    \label{fig:framework}
\end{figure*}
We focus the treatment on first-order processes for simplicity, although the approach is general and can be extended straightforwardly to second-order processes.
To describe the proposed approach, we start by recalling that the intensity of the peaks in the case of single-phonon resonant Raman processes can be calculated via the following relation~\cite{ganguly_theory_1967,yu_fundamentals_2010}:
\begin{equation}
    \label{eq:raman_main}
    I(\omega_{RS}) \propto 
    \sum_{\nu} 
    \left| 
    \mathbf{P}_{s} \, 
    \mathbf{R}_{\nu}(E_{L}) \,
    \mathbf{P}_{i}
    \right|^2 
    \delta(\omega_{RS} \pm  \omega_{\nu})
    ~,
\end{equation}
where $I(\omega_{RS})$ is the Raman intensity at Raman-shift frequency $\omega_{RS}$, $\mathbf{R}_{\nu}(E_L)$ is the mode-resolved Raman tensor for phonon mode $\nu$ evaluated at the incident laser energy $E_L$, $\mathbf{P}_i$ and $\mathbf{P}_s$ are the polarization unit vectors of the incident and scattered light, respectively.
The central quantity in Eq.~\eqref{eq:raman_main} is the Raman tensor, which can be evaluated using the expression from third-order perturbation theory~\cite{venezuela_theory_2011}:
\begin{multline}
    \label{eq:raman_tensor}
    \mathbf{R}_{\nu}(E_{L}) = 
    \sum_{\mathbf{k}} \sum_{f, n, n'} \mathbf{D}_{fn'}(\mathbf{k}) \,
    g_{n'n\nu}(\mathbf{k},\mathbf{q}=\Gamma) \,
    \mathbf{D}_{nf}(\mathbf{k}) \times\\
    \frac{
    1
    }{
    (E_{L} - \Delta\varepsilon_{nf} - i \gamma_{n})
    (E_{L} - \Delta\varepsilon_{n'f} \pm \hbar \omega_{\nu} - i \gamma_{n'})
    }
    ~,
\end{multline}
where $\mathbf{D}_{nm}(\mathbf{k})$ are optical-dipole matrix elements between electronic states $n$ and $m$ at momentum $\mathbf{k}$,
$g_{n'n\nu}(\mathbf{k},\mathbf{q}=\Gamma)$ is the electron-phonon coupling matrix element connecting electronic states $n$ and $n'$ via phonon mode $\nu$ at phonon wavevector $\mathbf{q}=\Gamma$, $\Delta\varepsilon_{nf}=\varepsilon_{n\mathbf{k}}-\varepsilon_{f\mathbf{k}}$, where $\varepsilon_{n\mathbf{k}}$ is the electronic energy level of band $n$ at momentum $\mathbf{k}$, $\gamma_n$ is the phenomenological linewidth of the intermediate electronic state $n$, and the $\pm\,\hbar\omega_\nu$ term accounts for Stokes and anti-Stokes processes, respectively.
As can be noticed, the calculation of the Raman tensor requires different ingredients that must be computed, very importantly, on fine k-point grids.
The main bottleneck to evaluate Eq.~\eqref{eq:raman_tensor} consistently at a level beyond (semi)local DFT is that $g_{mn\nu}$ are not readily accessible.
Linear-response techniques that can calculate $g_{mn\nu}$ have been derived only for a few selected methods, notably density-functional perturbation theory (DFPT) for (semi)local DFT and DFT$+U$~\cite{hubbard_ep_2021,paper_yang_first-principles_2025} and, more recently, $GW$ perturbation theory (GWPT)~\cite{li_electron-phonon_2019,li_electron-phonon_2024}.
The latter, however, is not yet broadly available in widely distributed public codes.
Moreover, other widely used methods such as hybrid XC functionals, meta-GGAs, or Koopmans-compliant functionals lack a DFPT or linear-response counterpart altogether.
To overcome this limitation, we propose a general route, based on a finite-difference (FD) approach, that allows one to calculate the full Raman tensor, including electron-phonon couplings, for any electronic structure method.
To illustrate the proposed approach to calculate first-order resonant Raman spectra, we schematically represent its main steps in Fig.~\ref{fig:framework}.
Assuming an already optimized crystal structure, a symmetry analysis is first performed to identify the inequivalent atomic displacements required in the following steps (Fig.~\ref{fig:framework}(a)).
Ground-state calculations are then carried out for the pristine structure and for each displaced configuration (Fig.~\ref{fig:framework}(b)), providing the electronic eigenvalues $\varepsilon_{n\mathbf{k}}$, the wavefunctions, and the forces acting on the atoms.
From the pristine calculation, the optical (dipole) matrix elements $\mathbf{D}_{nm}(\mathbf{k})$ can be readily obtained~\cite{lee_electronphonon_2023} (Fig.~\ref{fig:framework}(d)).
The remaining ingredients of Eq.~\eqref{eq:raman_tensor}, namely the phonons and the electron-phonon couplings, are, as already mentioned, the most challenging quantities to obtain.
Here, we suggest calculating them from the ground-state calculations of the displaced configurations.
The forces of all these configurations are employed to calculate the phonon frequencies and eigenvectors (Fig.~\ref{fig:framework}(c)), as already routinely performed in the literature~\cite{phonopy_2015, Togo_2023, bastonero_automated_2024}.
While finite-difference techniques for phonons are well established, extending them to electron-phonon matrix elements is far less straightforward.
The key issue in the calculation of $g_{mn\nu}$ is the evaluation of the matrix element of the change of the self-consistent effective potential with respect to the atomic displacements $\tau_{\kappa\alpha l}$:
\begin{equation}
    \bra{ \psi_{m\textbf{k}+\textbf{q}}} \dfrac{\partial V}{\partial \tau_{\kappa\alpha l}} \ket{{\psi_{n\textbf{k}}}}
    ,
    \label{eq:dVdt}
\end{equation}
where $\tau_{\kappa\alpha l} = \mathbf{R}_l + \tau_{\kappa\alpha}$ is the $\alpha$ Cartesian component of the position of the atom $\kappa$ in a unit cell located at position $\mathbf{R}_l$, with the index $l$ running over the unit cells.
In this work, we calculate these matrix elements by extending the eigenvalue-projectability approach introduced in Ref.~\citenum{poliukhin_carrier_2025} to the case of finite-$\textbf{k}$ sampling of displaced cells at $\textbf{q}=\Gamma$ (Fig.~\ref{fig:framework}(e)):
\begin{multline}
    \bra{\psi_{m\textbf{k}}} \dfrac{\partial V}{\partial \tau_{\kappa\alpha l}} \ket{\psi_{n\textbf{k}}} =\\ \frac{\tau^{-1}_{\kappa\alpha l}}{2}\Big(\sum_{j} \varepsilon^+_{j\textbf{k}} u_{m\textbf{k}j}^{*+}  u^{+}_{jn \textbf{k}}   
    -  \sum_{j} \varepsilon^-_{j\textbf{k}} u_{m\textbf{k}j }^{*-} u^{-}_{jn \textbf{k}}\Big)
    ~,
    \label{eq:projectability}
\end{multline}
where
\begin{equation}
    u^{\pm}_{jn\textbf{k}} = \braket{\psi^\pm_{j\textbf{k}} | \psi_{n\textbf{k}}}
    ~;
    \label{eq:projectability_u} 
\end{equation}
here, $\psi^\pm_{n\textbf{k}}$ and $\varepsilon^\pm_{n\textbf{k}}$ are the wavefunctions and eigenvalues of the perturbed cells, and the $+$ and $-$ symbols denote positive and negative atomic displacements along Cartesian directions, respectively.
The advantage of this expression is that the electron-phonon matrix elements can be obtained from quantities any electronic-structure code already stores at the end of a total-energy calculation.
This means that any method able to calculate forces and an effective Hamiltonian can, in principle, be employed.
A further practical advantage of Eq.~\ref{eq:projectability} over the general $\mathbf{q}\neq\Gamma$ expression is that the overlaps in Eq.~\ref{eq:projectability_u} are evaluated in the original unit cell, without the need to unfold the wavefunctions to a supercell.
This follows from the fact that first-order Raman scattering only involves $\mathbf{q}=\Gamma$ phonons, which preserve the periodicity of the unit cell.
In practice, the most demanding step in the evaluation of Eq.~\ref{eq:projectability} is the calculation of the overlaps between displaced and pristine wavefunctions on the coarse k-grid.
Fortunately, this cost is greatly reduced using the point-group symmetries of the displaced configurations, which reduce the number of calculations from $6N_{\mathrm{at}}$ to $N_{\mathrm{ineq}}$, where $N_{\mathrm{ineq}}$ is the number of inequivalent displacements (we refer the reader to Ref.~\citenum{poliukhin_carrier_2025} for the full details of this symmetry-reduction procedure).
The last step before calculating the Raman tensor, represented in Fig.~\ref{fig:framework}(f), is to interpolate into a fine k-point mesh the electronic energies $\varepsilon_{n\mathbf{k}}$, the dipole moments $\mathbf{D}_{nm}(\mathbf{k})$, and the electron-phonon couplings $g_{nm\nu}(\mathbf{k)}$ with Wannier interpolations~\cite{marzari_maximally_1997, marzari_maximally_2012, Giustino_2007}.
Interpolations avoid performing expensive self-consistent calculations directly on such fine meshes; in fact, realistic spectra may require grids of the order of $150\times 150\times 1$ k-points, as in graphene, where a direct calculation would be computationally prohibitive.
Another key difference in this work with respect to previous studies~\cite{Marini2024, huang_qr2-code_2025} is the use of a robust Wannierization protocol based on the projectability-disentanglement~\cite{qiao_projectability_2023}, bypassing a manual, and material-specific procedure.
Finally, interpolated quantities are used to calculate the Raman tensor at a given laser frequency and light polarizations, as shown in Fig.~\ref{fig:framework}(g).
The full workflow is benchmarked against DFPT at the PBE level and reported in the Supporting Information~\cite{supplement}, where the FD and DFPT spectra of graphene and MoS$_2$ are in quantitative agreement.

Thanks to this formulation, we can finally explore the effects of different electronic structure methods on the resonant Raman spectra of two-dimensional (2D) materials.
In fact, the reduced dielectric screening of these materials makes the electronic band structure and electron-phonon matrix elements -- both required for the Raman tensor Eq.~\eqref{eq:raman_tensor} -- very sensitive to the level of theory~\cite{qiu_screening_2016}. 
We will not consider excitonic and other many-body effects which, although known to be important in these materials (e.g., see Refs.~\citenum{reichardt_nonadiabatic_2020,Guandalini2025}), require also the Bethe-Salpeter equation.\nocite{Guandalini2025}
As prototypical examples, we select graphene and monolayer MoS$_2$: graphene is a semimetal that is in resonance at any laser energy and exhibits a single first-order Raman-active mode (G), while MoS$_2$ is a 2D semiconductor, with two first-order Raman-active modes ($E'$ and $A'_1$) and a finite gap that selects the laser energies at which the resonance condition is met.
For each material, we compute Raman spectra with three different XC functionals, namely semilocal PBE~\cite{Perdew1996a}, meta-GGA R$^2$SCAN~\cite{Furness2020}, and hybrid HSE06~\cite{Heyd2003,Heyd2006}; for MoS$_2$, we also compute the spectra using one-shot $G_0W_0$~\cite{gw_1985,marini_yambo_2009}.
In the following, we systematically compare the results obtained with the different methods against experiment.

\begin{figure}
    \centering
    \includegraphics[width=0.96\linewidth]{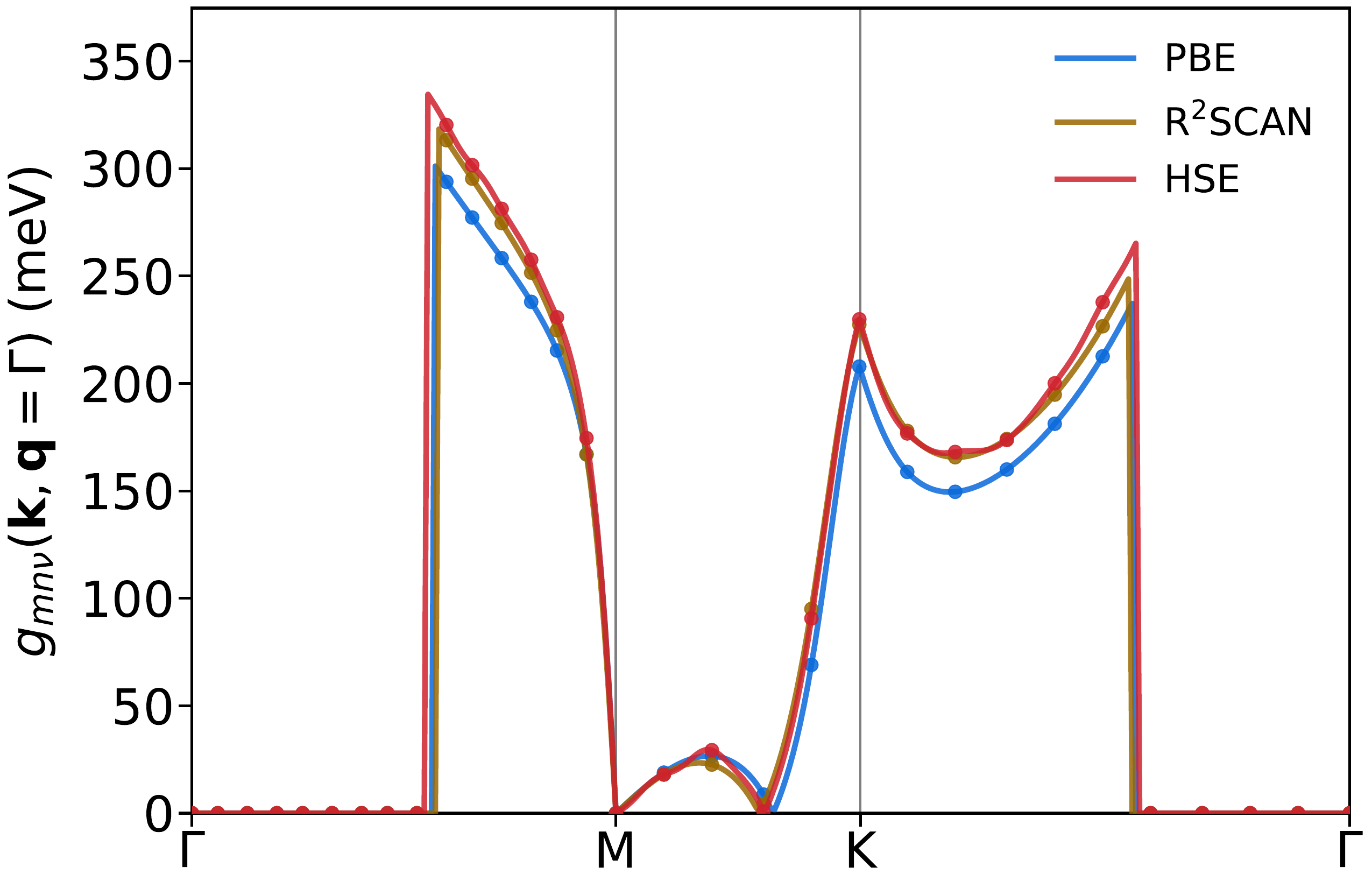}
    \caption{
    \label{fig:epc_graphene} 
    Electron-phonon matrix elements of graphene computed with different methods along the $\mathbf{\Gamma}$-$\mathbf{M}$-$\mathbf{K}$-$\mathbf{\Gamma}$ path at $\mathbf{q}=\Gamma$, for the electronic states $m=2$, $n=3$ coupled by the Raman-active G mode ($\nu=6$). All quantities are interpolated with EPW from a coarse $30\times 30\times 1$ \textbf{k}-grid; circles mark the values obtained directly on that grid.
    }
\end{figure}

Starting with graphene, this material has only one Raman-active phonon mode (the G mode at $\textbf{q}=\Gamma$); therefore, the most informative quantity to compare with experiment is the ratio of Stokes and anti-Stokes peak intensities at different laser energies.
In fact, this ratio probes simultaneously the strength of the electron-phonon coupling and the energy dependence of the resonance.
\begin{figure}
\includegraphics[width=0.96\linewidth]{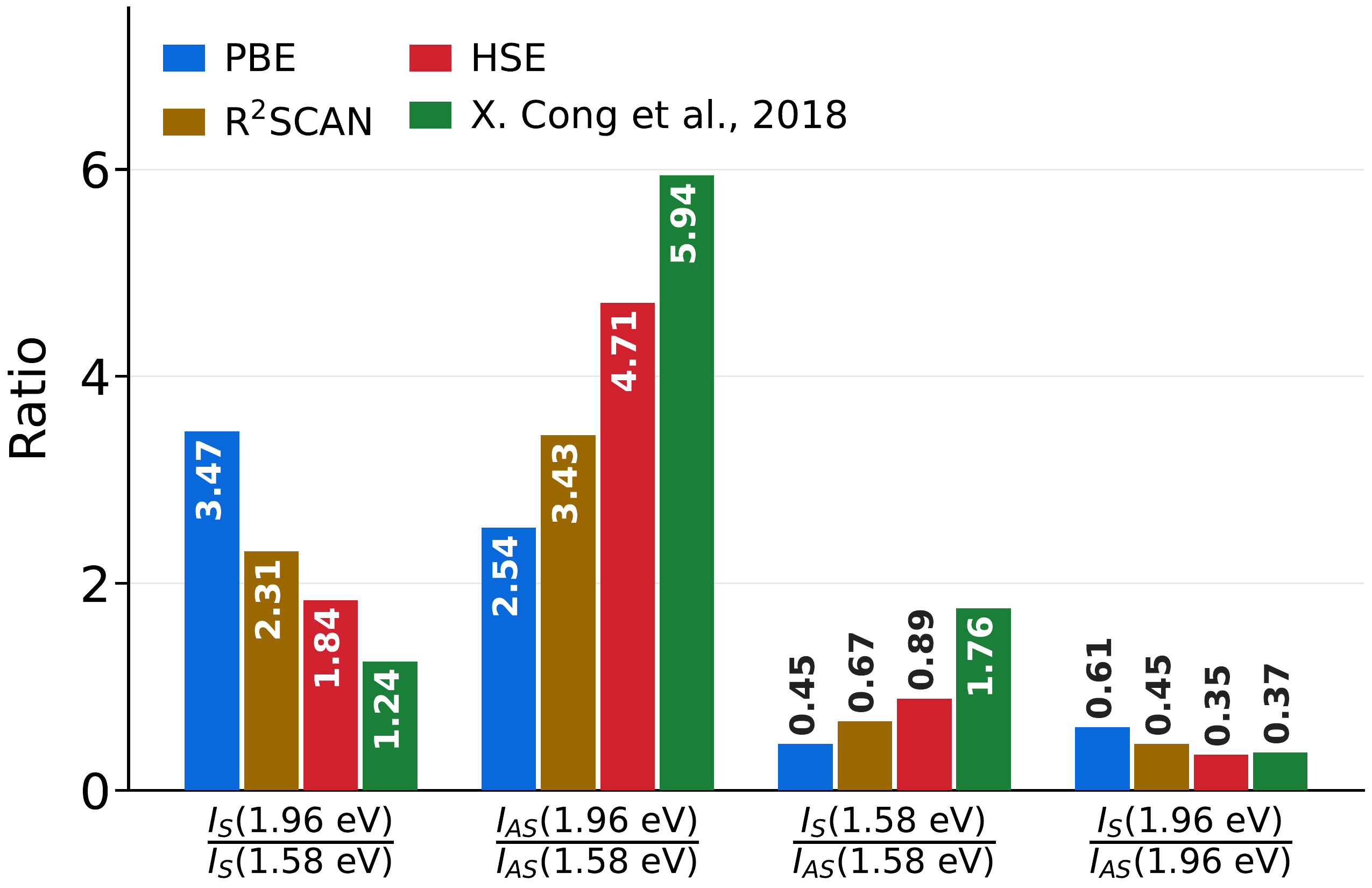}
    \caption{\label{fig:graphene_decomp} Ratios of Raman intensities of the G mode of graphene computed with different electronic-structure methods at laser energies of 1.58~eV and 1.96~eV. Stokes (S)/anti-Stokes (AS) ratios are reported, to enable direct comparison with experiment~\cite{cong_stokes_2018}.
}
\end{figure}
We consider two laser energies of 1.58~eV and 1.96~eV, and report all possible combinations of Stokes ($I_S$) and/or anti-Stokes ($I_{AS}$) intensity ratios in Fig.~\ref{fig:graphene_decomp}, for all the functionals considered.
For all four ratios considered here, HSE06 provides the best agreement with experiment, while R$^2$SCAN lies systematically between HSE06 and PBE.
It is of interest to understand which quantities in Eq.~\eqref{eq:raman_tensor}, namely electronic eigenvalues, electron-phonon matrix elements, and phonon frequencies, have a major impact on the intensity ratios and from which functional one should pick the values.
To this end, we select PBE quantities as a baseline, and substitute one of them in Eq.~\eqref{eq:raman_tensor} with its XC-calculated counterpart, while keeping the other two at the PBE level; we denote the resulting calculations $\mathrm{XC}[\varepsilon_{n\mathbf{k}}]$, $\mathrm{XC}[g_{mn\nu}(\mathbf{k})]$, and $\mathrm{XC}[\omega_{G}]$, according to which ingredient has been substituted.
To assess the overall performance of each method across the four ratios at once, rather than column by column, we employ a single aggregate score, the mean absolute log-ratio error, $\mathrm{MALE} = \tfrac{1}{4}\sum_{i=1}^{4} \left| \ln\left( R_i^{\rm theo}/R_i^{\rm exp} \right) \right|$, where $R_i^{\rm theo}$ and $R_i^{\rm exp}$ are the calculated and experimental values of the $i$-th ratio in Tab.~\ref{tb:ratios_graphene}.
We adopt this log-space definition, rather than a conventional relative error, because it weighs over- and under-estimation symmetrically, so that a twofold overestimate and a twofold underestimate contribute equally. 
This offers a robust summary of how consistently each method reproduces the four measured ratios.
\begin{table*}

\begin{ruledtabular}
\small
\setlength{\tabcolsep}{4pt}
\begin{tabular}{ccccccc}
Method & $\frac{I_S(1.96\ \mathrm{eV})}{I_S(1.58\ \mathrm{eV})}$ & $\frac{I_{AS}(1.96\ \mathrm{eV})}{I_{AS}(1.58\ \mathrm{eV})}$ & $\frac{I_S(1.58\ \mathrm{eV})}{I_{AS}(1.58\ \mathrm{eV})}$ & $\frac{I_S(1.96\ \mathrm{eV})}{I_{AS}(1.96\ \mathrm{eV})}$ & MALE & $\omega_G\ (\mathrm{cm}^{-1})$ \\
\hline
Experiment~\cite{cong_stokes_2018} & 1.24 & 5.94 & 1.76 & 0.37 & -- & 1580 \\
\hline
HSE & {1.84} & 4.71 & 0.89 & \textbf{0.35} & \textbf{0.34} & 1544 \\
R$^2$SCAN & 2.31 & 3.43 & 0.67 & 0.45 & 0.58 & \textbf{1545} \\
PBE & 3.47 & 2.54 & 0.45 & 0.61 & 0.94 & 1538 \\
\hline
HSE[$\varepsilon_{n\mathbf{k}}$] & 4.48 & \textbf{6.21} & \textbf{1.05} & 0.76 & 0.64 & -- \\
R$^2$SCAN[$\varepsilon_{n\mathbf{k}}$] & 2.55 & 2.63 & 0.59 & 0.57 & 0.77 & -- \\
\hline
HSE[$g_{mn\nu}(\mathbf{k})$] & \textbf{1.65} & 0.94 & 0.60 & {1.05} & 1.06 & -- \\
R$^2$SCAN[$g_{mn\nu}(\mathbf{k})$] & 6.75 & 4.20 & 0.35 & 0.56 & 1.02 & -- \\
\hline
HSE[$\omega_{G}$] & 3.48 & 2.55 & 0.45 & 0.61 & 0.94 & -- \\
R$^2$SCAN[$\omega_{G}$] & 3.48 & 2.58 & 0.44 & 0.60 & 0.93 & --
\end{tabular}
\end{ruledtabular}

\caption{Ratios of intensities of the G peak of graphene for the Stokes and anti-Stokes process with different functionals at laser energies 1.58 eV and 1.96 eV, together with the peak position of the G band. The labels XC[$\varepsilon_{n\mathbf{k}}$], XC[$g_{mn\nu}(\mathbf{k})$], and XC[$\omega_{G}$] denote calculations where only a single quantity is computed with the ``XC'' functional, while the other two are kept at the PBE level: the electronic eigenvalues, the electron-phonon matrix elements, or the phonon frequency, respectively. For each column, the value in closest agreement with experiment is highlighted in bold. MALE is the mean absolute log-ratio error with respect to experiment over the four intensity-ratio columns, $\mathrm{MALE} = 1/4\sum_{i=1}^{4} \left| \ln\left( R_i^{\rm theo}/R_i^{\rm exp} \right) \right|$, where $R$ stands for intensity ratio; the lowest (best) value is highlighted in bold.
}
\label{tb:ratios_graphene}
\end{table*}
The corresponding ratios are reported in Tab.~\ref{tb:ratios_graphene}.
By comparing the results with respect to PBE and experiments, the decomposition makes it clear that the electronic eigenvalues, which enter the resonant denominators, and the electron-phonon couplings are the two factors that dominate the Raman ratios.
Substituting only the phonon frequencies has a marginal effect (we note that, in this case, the frequencies from the different XC functionals are rather similar; this effect may be greater in other materials, such as transition-metal oxides).
R$^2$SCAN, despite being a meta-GGA functional, yields electronic eigenvalues closer to PBE than to HSE, so the resonance denominators shift only mildly. As a result, it gives an intermediate description of the Raman ratios, improving over PBE while remaining less accurate than HSE.
We further note that the Raman tensor depends on the square of the electron-phonon matrix elements.
This means that even the small differences in electron-phonon couplings are amplified into larger deviations, as seen in Tab.~\ref{tb:ratios_graphene}.
The aggregate MALE score in Tab.~\ref{tb:ratios_graphene} confirms this picture quantitatively: the fully consistent HSE calculation achieves the lowest error (MALE~=~0.34), clearly outperforming every partial substitution, including HSE[$\varepsilon_{n\mathbf{k}}$] (0.64) and HSE[$g_{mn\nu}(\mathbf{k})$] (1.06) -- the latter even worse than plain PBE (0.94).
While mixing HSE-computed quantities may occasionally improve agreement on an individual ratio, as seen in Tab.~\ref{tb:ratios_graphene}, no such partial substitution matches the accuracy of the fully consistent calculation.
The systematic enhancement of the electron-phonon couplings with more advanced electronic-structure methods, at the origin of better predictions, is most directly seen by examining them for graphene along a high-symmetry path of the Brillouin zone, shown in Fig.~\ref{fig:epc_graphene} for band indices $m=2$, $n=3$ and the G phonon mode ($\nu=6$).
We note that both the R$^2$SCAN and HSE functionals produce a similar enhancement of the electron-phonon couplings with respect to PBE, of the order of 5--10\% along the path, in good agreement with previous studies~\cite{zhou_abinitio_2021,li_electron-phonon_2024,wang_accurate_2024,abramovitch_electron-phonon_2025}.
This trend can be rationalized by the reduction of the dielectric overscreening present in semilocal DFT~\cite{yin_corelated_2013}: since the electron-phonon matrix elements can be expressed as the matrix elements of the variation of the bare external potential screened by the inverse dielectric matrix~\cite{Giustino_2017}, a smaller dielectric response yields larger couplings.
While this trend is general, it is most pronounced in insulators, where DFT severely underestimates the band gap and consequently overestimates the static dielectric constant, as we will see later.
Finally, all approaches reproduce the position of the G band within 5\%, with HSE and R$^2$SCAN yielding nearly identical frequencies (1544 and 1545~cm$^{-1}$, respectively) and PBE showing the largest error.
This is consistent with the established observation that peak positions are already captured well at the DFT level, whereas intensities are not.

As a second example, we turn now to MoS$_2$, where at $\mathbf{q}=\Gamma$ the monolayer has two Raman-active modes, the in-plane $E'$ and the out-of-plane $A'_1$, whose relative intensity at a given laser energy is accessible experimentally~\cite{placidi_multiwavelength_2015}.
We calculate and compare the spectra for a laser energy of 3.81~eV, above the optical gap, in order to minimize the contribution of the exciton-phonon coupling, which becomes important at lower (excitonic) energies~\cite{reichardt_nonadiabatic_2020} and which, as already discussed, lies outside the scope of the approximation adopted here.
To further demonstrate the ability of the proposed framework to handle any electronic structure method, we calculate the resonant Raman spectra using also many-body perturbation theory through the one-shot $GW$ ($G_0W_0$) approximation~\cite{marini_yambo_2009}, in addition to the PBE and HSE results; such calculations are accessible in this finite-gap material, whereas they are computationally much more delicate in semimetals such as graphene~\cite{guandalini_efficient_2024}.
We exclude R$^2$SCAN calculations for \ce{MoS2} due to convergence issues of the forces and phonons described in the Supporting Information~\cite{supplement}.
\begin{figure}
    \centering
    \includegraphics[width=0.96\linewidth]{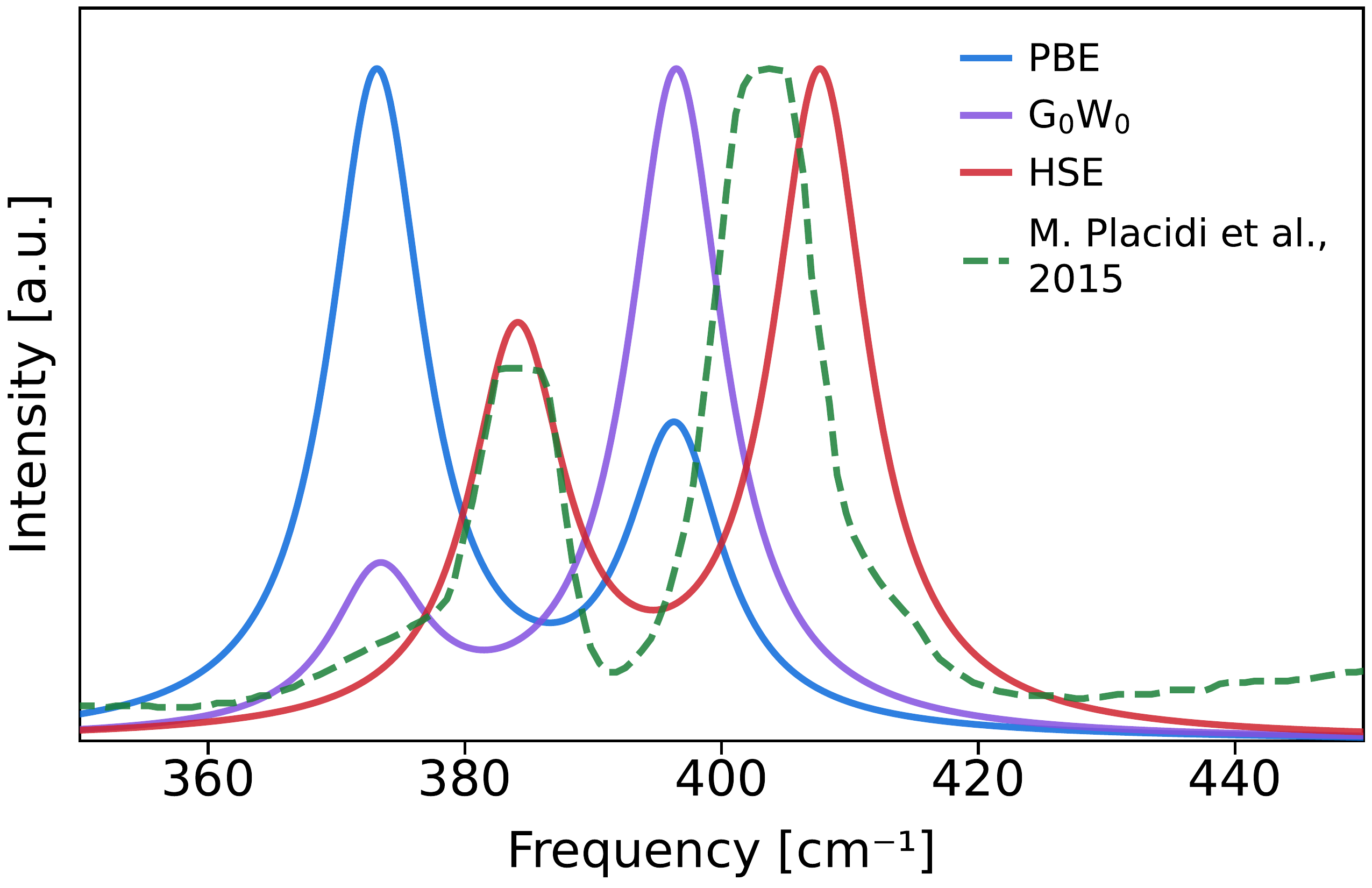}
    \caption{\label{fig:mos2_int} Comparison of the Raman spectrum of MoS$_2$ monolayer computed with different electronic structure methods and compared to experiment from Ref.~\citenum{placidi_multiwavelength_2015} at a laser energy of 3.81~eV. Theoretical intensities are smeared with a 5~cm$^{-1}$ wide (FWHM) Lorentzian.
}
\end{figure}
The Raman spectra calculated with the different methods are compared to experiments~\cite{placidi_multiwavelength_2015} in Fig.~\ref{fig:mos2_int}.
At the PBE level the two peaks are both quantitatively and qualitatively wrong, with the predicted intensity ordering $I_{A'_1}<I_{E'}$ being inverted with respect to the experimental reference.
Both $G_0W_0$ and HSE recover the correct ordering; HSE further demonstrates excellent agreement with respect to the experimental intensity ratio.
HSE also improves the peak positions, shifting them from 373 and 396~cm$^{-1}$ at the PBE level to 385 and 408~cm$^{-1}$, in close agreement with the experimental values of 386 and 404~cm$^{-1}$~\cite{placidi_multiwavelength_2015}.
\begin{table}

    \begin{ruledtabular}
    \begin{tabular}{ccccc}
        Method & $I_{A^{'}_{1}}/I_{E^{'}}$ & $E_{g}$, eV & $\omega_{E'}$ (cm$^{-1}$) &   $\omega_{A'_1}$ (cm$^{-1}$)\\ 
        \hline
        Experiment & 1.8 & 1.9 & 386 & 404 \\
        \hline
        HSE & \textbf{1.7} & \textbf{2.1} & \textbf{385} & \textbf{408} \\
        G$_0$W$_0$ & 4.2 & 2.5 & -- & -- \\
        PBE & 0.5 & 1.7 & 373 & 396 \\
        \hline
        HSE[$\varepsilon_{n\mathbf{k}}$] & 2.1 & -- & -- & --\\
        G$_0$W$_0$[$\varepsilon_{n\mathbf{k}}$] & 3.0 & -- & -- & --
    \end{tabular}
    \end{ruledtabular}
        \caption{Ratios between A$^{'}_1$ and E$^{'}$ Raman-active peaks of 1-layer MoS$_2$ at laser energy of 3.81 eV calculated with different functionals. To emphasize the importance of correctly predicting the band structure, the band gap is also reported. The notation XC[$\varepsilon_{n\mathbf{k}}$] follows Tab.~\ref{tb:ratios_graphene}: only the electronic eigenvalues are computed with the ``XC'' functional, while all other quantities are kept at the PBE level. For each column, the value in closest agreement with experiment is highlighted in bold.
    }
    \label{tb:ratios_mos2}
\end{table}
Unlike graphene, MoS$_2$ has a finite band gap that beyond-DFT methods generally widen with respect to the (semi)local DFT value, due to the well-known band-gap underestimation of the latter~\cite{perdew_physical_1983}.
Since the resonance condition in Eq.~\eqref{eq:raman_tensor} depends sensitively on the energy denominators, even modest changes in the band gap can substantially redistribute the Raman intensities.
As for graphene, it is interesting to disentangle the role of the band gap from that of the electron-phonon coupling.
To do so, we adopt the same substitution scheme introduced above: starting from the PBE baseline, we substitute only the electronic eigenvalues with their HSE or $G_0W_0$ counterparts, denoted $\mathrm{XC}[\varepsilon_{n\mathbf{k}}]$ in Tab.~\ref{tb:ratios_mos2}, while keeping the rest at the PBE level.
Tab.~\ref{tb:ratios_mos2} reports this comparison alongside the PBE, HSE, and $G_0W_0$ results, where all quantities are computed at the same, respective level of theory.
In both the HSE and $G_0W_0$ cases, the substitution of only the electronic band energies results in a significantly better intensity ratio than plain PBE.
Since both HSE and $G_0W_0$ do not particularly alter the shape of the electronic eigenvalues, the primary cause of the improvement resides in the better estimation of the band gap.
For this reason, HSE substitutions outperform $G_0W_0$ thanks to better agreement with the experimental band gap.
Notably, the two methods behave differently once the electron-phonon couplings are also computed at the same level of theory: the HSE ratio improves further, from 2.1 to 1.7, whereas the $G_0W_0$ ratio worsens, from 3.0 to 4.2.
Although the enhancement of the electron-phonon matrix elements near the $\mathbf{K}$ point is very similar for HSE and $G_0W_0$, the $G_0W_0$ correction is less uniform at other $\mathbf{k}$ points along the high-symmetry path (see Supporting Information~\cite{supplement}); by contrast, HSE produces a more uniform enhancement of the coupling, which combined with its closer-to-experiment band gap explains the markedly better agreement of HSE with the measured Raman ratio.
The fact that HSE outperforms one-shot $G_0W_0$ further suggests that a self-consistent $GW$ scheme, which would partially correct the gap overestimation, would likely also improve the agreement with experiment for the Raman ratios.
Another possible explanation for the $G_0W_0$ underperformance is the slight inconsistency in the calculation of the electron-phonon couplings.
In fact, the proposed approach uses wavefunction projections to calculate electron-phonon matrix elements (see Eq.~\eqref{eq:projectability}), and, since $G_0W_0$ does not provide updated wavefunctions, but only electronic band energies, the calculation combines quantities that are not fully consistent with each other.
As already noted for graphene, a consistent level of theory also proves essential for \ce{MoS2} in order to maximize the agreement with experiment.

In summary, we presented an efficient and generally applicable approach for calculating resonant Raman spectra in crystal structures for any electronic-structure method, also unrelated to DFT, overcoming the common limitations of semilocal DFT.
We demonstrated the capabilities of this approach on two selected 2D materials using different XC functionals and many-body perturbation theory.
Among the methods tested, HSE offers the best overall accuracy on the spectral intensities, while performing at least as well as the other methods on the peak positions.
Notably, we showed that accurate predictions are obtained when a consistent level of theory is employed in the calculation of the Raman tensor.
As a consequence, replacing only the electronic eigenvalues or only the electron-phonon coupling with their beyond-DFT counterparts, while leaving the rest of the calculation at the DFT level, can move the Raman ratios further from experiment than a plain DFT calculation.
As the proposed workflow needs only forces, eigenvalues, and wavefunctions at displaced configurations, it applies directly to other 2D materials and to the growing databases of computed Raman spectra~\cite{Taghizadeh2020}, opening the door to systematic, beyond-DFT Raman libraries for benchmarking against experiment.

\section*{Data availability}
The data needed to reproduce the results of this work can be found in the Materials Cloud Archive~\cite{talirz_materials_2020}.

\section*{Code availability}
The main source code for the calculation of electron-phonon couplings with arbitrary electronic structure method is available on GitHub (https://github.com/Koulb/ElePhAny.jl)~\cite{poliukhin_carrier_2025}.
Ground-state calculations are performed with \textsc{Quantum ESPRESSO}~\cite{Giannozzi_2009,Giannozzi_2017}, phonons with \texttt{Phonopy}~\cite{phonopy_2015}, Wannier interpolation with \texttt{EPW}~\cite{lee_electronphonon_2023}, and the final assembly of the Raman tensor with \texttt{QR$^2$}~\cite{huang_qr2-code_2025}, that are all open-source software.

\section*{Acknowledgements}
A.P. thanks Miki Bonacci for fruitful discussions about GW calculations.
The authors acknowledge support from the Swiss National Science Foundation, grant number 213082: ``Accurate and efficient electronic structure functionals for energies and spectra of materials''.
L.B. and N.M. gratefully acknowledge support from the Deutsche Forschungsgemeinschaft (DFG) under Germany’s Excellence Strategy (EXC 2077, No. 390741603, University Allowance, University of Bremen) and Lucio Colombi Ciacchi, the host of the ``U Bremen Excellence Chair Program''.
N.M. acknowledges support by the NCCR MARVEL, a National Centres of Competence in Research, funded by the Swiss National Science Foundation (grant number 205602).
The computational time has been provided by the Swiss National Supercomputing Centre (CSCS) under project ID mr33.
%


\section*{Competing interests}
The authors declare no competing interests.
%

\section*{References}
\putbib[main]
\end{bibunit}

\clearpage


\clearpage

\setcounter{equation}{0}
\renewcommand{\theequation}{S\arabic{equation}}
\setcounter{figure}{0}
\renewcommand{\thefigure}{S\arabic{figure}}
\setcounter{table}{0}
\renewcommand{\thetable}{S\arabic{table}}
\setcounter{affil}{0}

\begin{bibunit}[apsrev4-2]

\title{Supplementary Information:\\
Resonant Raman spectroscopies beyond density-functional theory}

\author{Aleksandr Poliukhin\orcidlink{0000-0001-8519-872X}}
\email{aleksandr.poliukhin@epfl.ch}
\affiliation{Theory and Simulation of Materials (THEOS), École polytechnique fédérale de Lausanne, 1015 Lausanne, Switzerland}
\author{Corto Babs Aubry}%
\affiliation{Theory and Simulation of Materials (THEOS), École polytechnique fédérale de Lausanne, 1015 Lausanne, Switzerland}
\author{Lorenzo Bastonero\orcidlink{0000-0001-9374-1876}}
\affiliation{U Bremen Excellence Chair, Bremen Center for Computational Materials Science, and MAPEX Center for Materials and Processes, University of Bremen, D-28359 Bremen, Germany}%
\author{Nicola Marzari\orcidlink{0000-0002-9764-0199}}
\affiliation{Theory and Simulation of Materials (THEOS), École polytechnique fédérale de Lausanne, 1015 Lausanne, Switzerland}
\affiliation{U Bremen Excellence Chair, Bremen Center for Computational Materials Science, and MAPEX Center for Materials and Processes, University of Bremen, D-28359 Bremen, Germany}
\affiliation{PSI Center for Scientific Computing, Theory and Data, 5232 Villigen PSI, Switzerland}%
\date{\today}

\maketitle
\onecolumngrid
\setcounter{section}{0}
\section{Comparison between DFPT and FD}
\begin{figure}[H]
    \centering
    \includegraphics[width=0.98\linewidth]{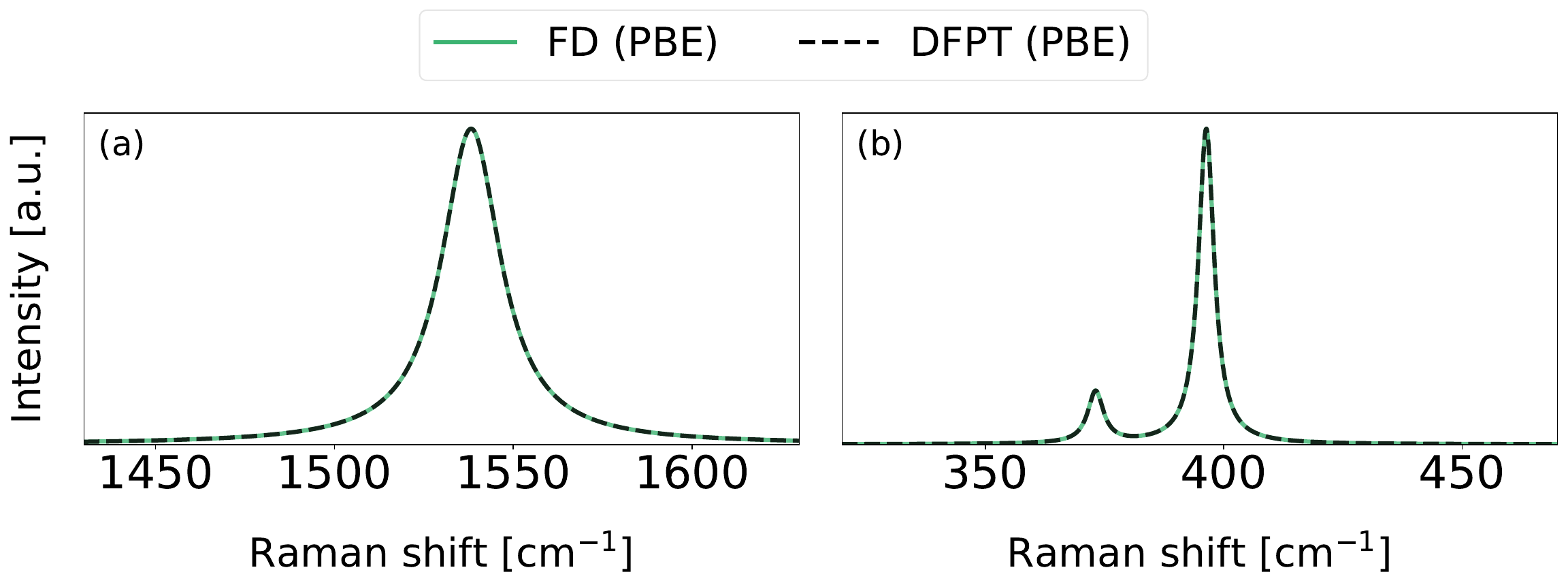}
    \caption{\label{fig:fd_dfpt_compare} Comparison of Raman intensities computed with FD and DFPT for (a) graphene at laser energy 1.96~eV and (b) MoS$_2$ at laser energy 2.41~eV. The close overlap of the two spectra shows that the electron-phonon coupling is consistent between FD and DFPT.}
\end{figure}
\section{Additional results for Graphene}
\begin{figure}[H]
    \centering
    \includegraphics[width=0.96\linewidth]{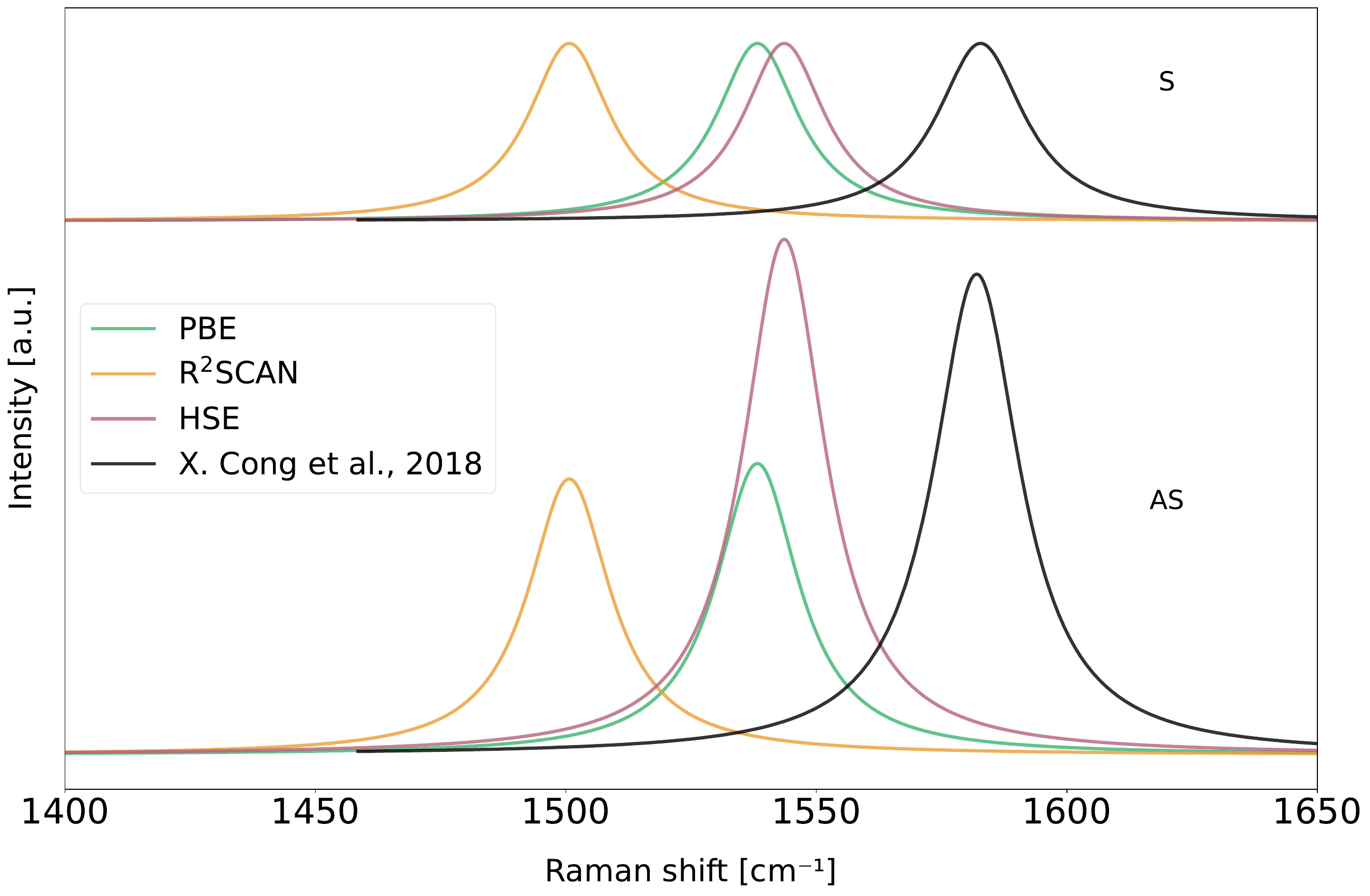}
    \caption{\label{fig:graphene_int} Comparison of the Stokes (S) and anti-Stokes (AS) Raman G peak of graphene computed with different methods and experiment~\cite{cong_stokes_2018} at laser energy 1.96 eV.}
\end{figure}

\begin{figure}[H]
    \centering
    \includegraphics[width=0.96\linewidth]{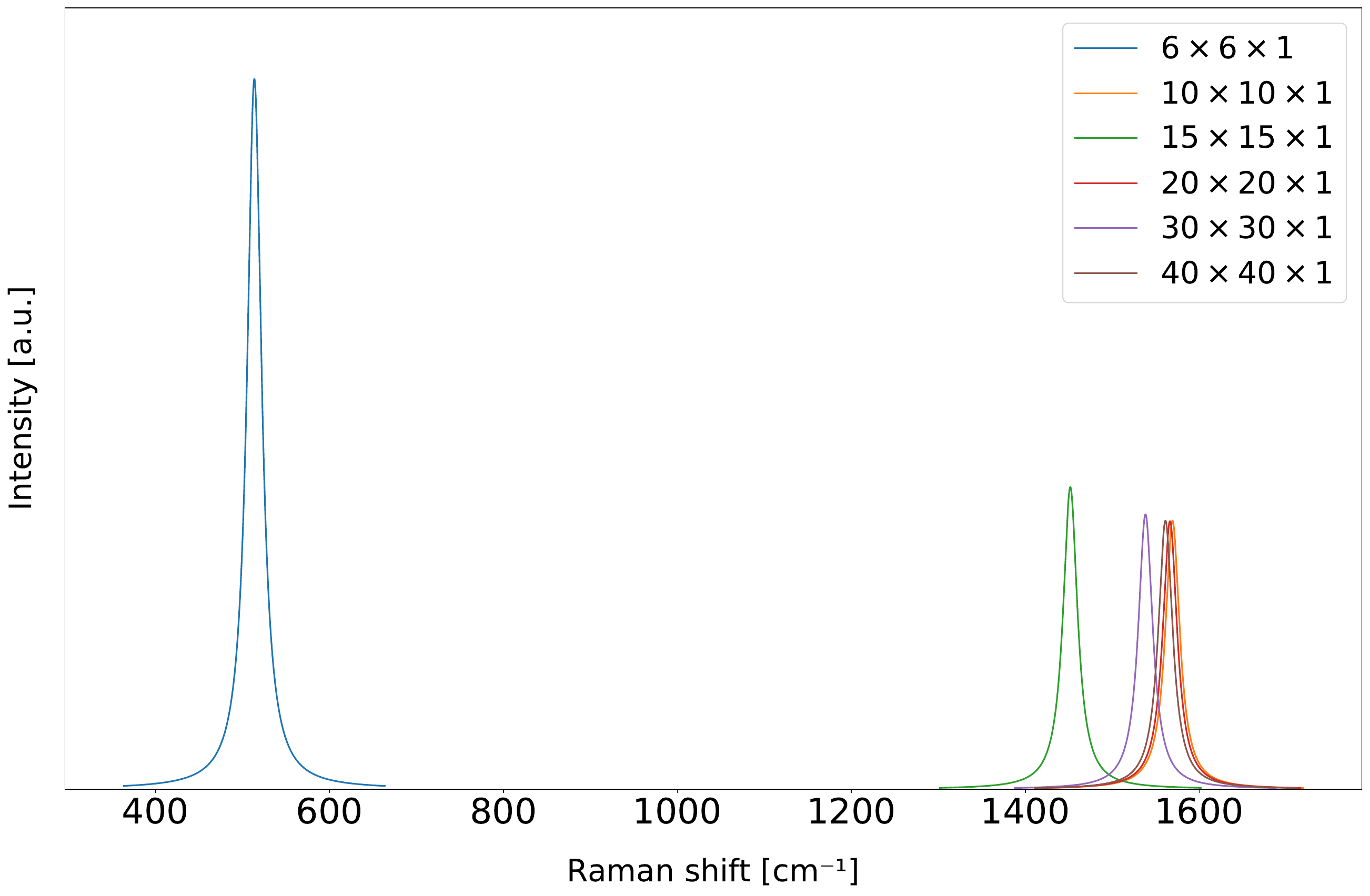}
    \caption{\label{fig:graphene_conv_int} Convergence of the DFPT Raman G peak of graphene for the Stokes process as a function of the coarse $\textbf{k}$ grid. For the final calculations we used the 30$\times$30$\times$1 grid that directly samples the $\textbf{K}$ point.}
\end{figure}
\section{Electron-Phonon matrix elements of MoS$_2$}
\begin{figure}[H]
\includegraphics[width=0.96\linewidth]{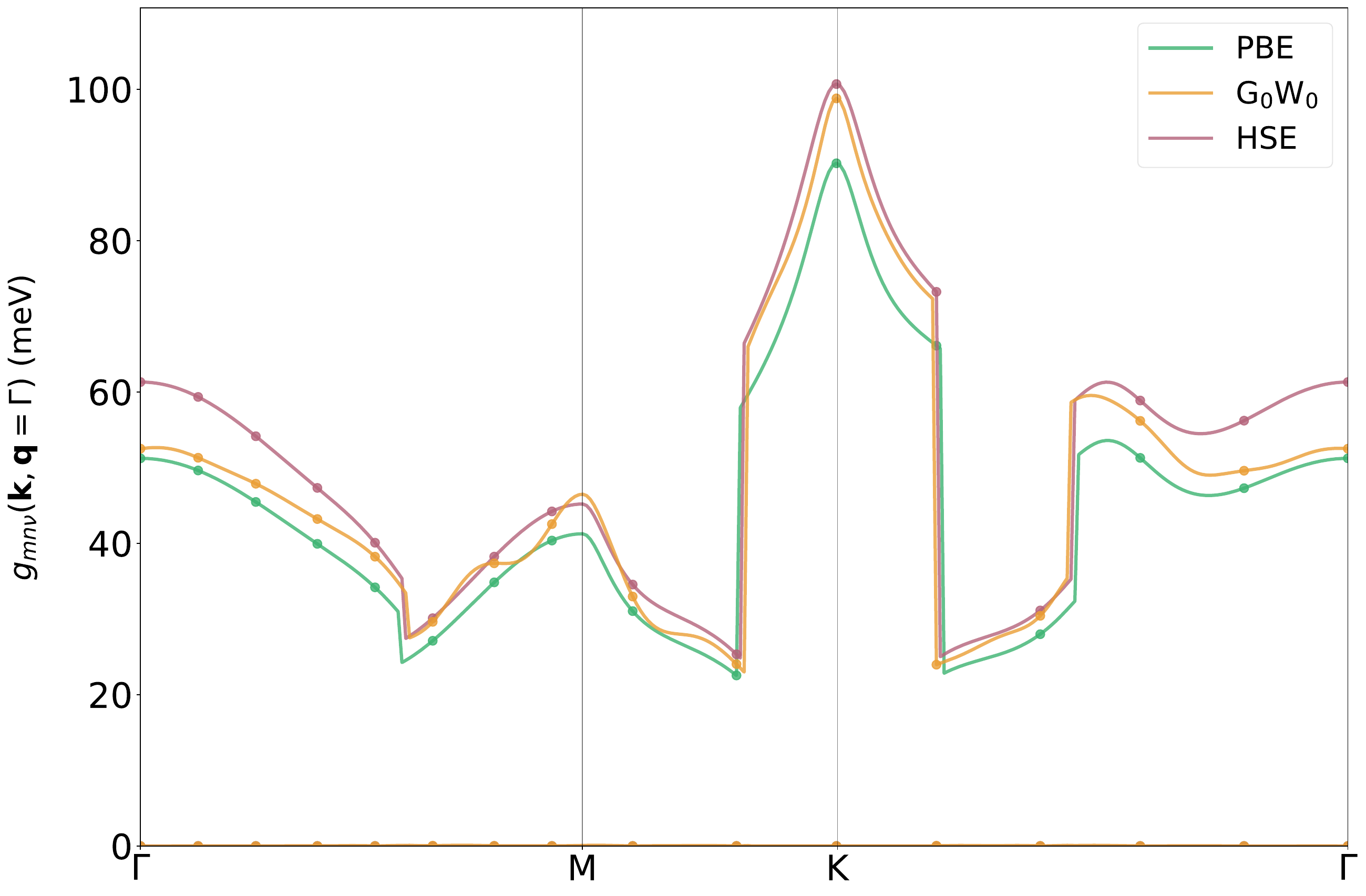}
    \caption{\label{fig:epc_mos2} Electron-phonon matrix elements of MoS$_2$ along a k-momentum path at q = $\Gamma$ for the electronic states ($m=2$, $n=7$) and phonon modes corresponding to E$^\prime$ and A$^\prime_1$ Raman-active modes. All the results are interpolated with EPW using a coarse 15×15×1 \textbf{k}-grid. The direct results obtained on the same 15×15×1 \textbf{k}-grid are highlighted with circles.
    }
\end{figure}

\section{Computational details}
We present here computational details for the calculations of graphene and MoS$_2$. 
For graphene, a $30\times 30\times 1$ coarse $\mathbf{k}$-point grid was used, chosen based on DFPT convergence tests of the Raman G peak (see Fig.~\ref{fig:graphene_conv_int}). 
A norm-conserving pseudopotential from the DOJO library~\cite{van_setten_pseudodojo_2018} was employed, with a kinetic energy cutoff (\texttt{ecut}) of 100~Ry and a convergence threshold of $1\times 10^{-13}$~Ry.
For the initial non-self-consistent field calculation, 20 bands were included. 
Wannier interpolation of the electron-phonon coupling and band energies was performed using 5 Wannier functions (\texttt{nbndsub}), constructed from $sp^2$ and $p_z$ projectors. 
Raman spectra were computed using a $150\times 150\times 1$ fine $\mathbf{k}$-point grid and a Lorentzian broadening of 20~cm$^{-1}$.
For the R$^2$SCAN calculation, the exchange-correlation functional was taken from the LIBXC library (IDs 497 and 498). 
For the HSE calculation, the default 0.25 fraction of exact exchange is used.

For MoS$_2$, a $15\times 15\times 1$ coarse $\mathbf{k}$-point grid was used. 
Norm-conserving pseudopotentials from the DOJO library were employed, with a kinetic energy cutoff of 100~Ry and a convergence threshold of $1\times 10^{-13}$~Ry.
For the initial non-self-consistent-field calculation, 17 bands were included without spin-orbit coupling. 
Wannier interpolation of the electron-phonon coupling and band energies was performed using 11 Wannier functions (\texttt{nbndsub}), where bands 1 to 6 were excluded from the Wannierization procedure. 
Projectors were chosen as angular momentum states with $l=2$ for the Mo atom and $l=1$ for the two S atoms.
Raman spectra were computed using a $150\times 150\times 1$ fine $\mathbf{k}$-point grid and a Lorentzian broadening of 4~cm$^{-1}$.
We experienced a complicated convergence trend with respect to \texttt{ecut} for forces and phonons with different flavors of SCAN and R$^2$SCAN functionals (LIBXC IDs 263, 267, 497, 498, 718, 719), even though the electron-phonon matrix elements were stable for even relatively small values of \texttt{ecut}. 
For this reason, we do not report the R$^2$SCAN results for MoS$_2$.
For the HSE calculation, the default 0.25 fraction of exact exchange is used.
For the $G_0W_0$ calculation, 400 states were used together with the plasmon pole approximation \cite{aryasetiawan_GW_1998} for the dielectric function evaluation, where a G-vector cutoff of 8~Ry for the dielectric matrix was used.
\putbib[si]
\end{bibunit}

\end{document}